\documentclass[preprint, prX]{revtex4}
\usepackage[all]{xy}
\usepackage{amssymb, amsmath}
\usepackage{accents}
\usepackage{graphicx}
\usepackage{subfig}

\begin{document}

\title{New families of superintegrable systems from Hermite and Laguerre exceptional orthogonal polynomials}
\author{ Ian Marquette}
\affiliation{School of Mathematics and Physics, The University of Queensland, Brisbane, QLD 4072, Australia}
\email{i.marquette@uq.edu.au,cquesne@ulb.ac.be}
\author{Christiane Quesne}
\affiliation{Physique Nucl\'eaire Th\'eorique et Physique Math\'ematique,
Universit\'e Libre de Bruxelles, Campus de la Plaine CP229, Boulevard du Triomphe, B-1050 Brussels, Belgium}
\email{i.marquette@uq.edu.au,cquesne@ulb.ac.be}

\begin{abstract}
In recent years, many exceptional orthogonal polynomials (EOP) were introduced and used to construct new families of 1D exactly solvable quantum potentials, some of which are shape invariant. In this paper, we construct from Hermite and Laguerre EOP and their related quantum systems new 2D superintegrable Hamiltonians with higher-order integrals of motion and the polynomial algebras generated by their integrals of motion. We obtain the finite-dimensional unitary representations of the polynomial algebras and the corresponding energy spectrum. We also point out a new type of degeneracies of the energy levels of these systems that is associated with holes in sequences of EOP. 
\end{abstract}
\maketitle

%%%%%%%%%%%%%%%%%%%%%%%%%%%%%%%%%%%%%%%%%%%%%%%%%%%%%%%%%
\section{Introduction}

In  classical mechanics, an $n$-dimensional Hamiltonian system $H$ with integrals of motion $X_{a}(\vec{x}, \vec{p})$ is called completely integrable (or Liouville integrable) if it possesses $n$ integrals of motion (including the Hamiltonian) that are well-defined functions on phase space, are in involution $\{H,X_{a}\}_{p}=0$, $\{X_{a},X_{b}\}_{p}=0$, $a,b=1,...,n-1$, and are functionally independent (where $\{,\}_{p}$ is the Poisson bracket). A
system is superintegrable if it is integrable and allows additional integrals of motion $Y_{b}(\vec{x},\vec{p})$, $\{H,Y_{b}\}_{p}=0$, $b=n,n+1,...,n+k$, that are also well-defined functions on phase space and such that the integrals $\{H,X_{1},...,X_{n-1},Y_{n},...,Y_{n+k}\}$ are functionally independent. A system is maximally superintegrable if the set contains $2n-1$ functions (i.e., $k=n-2$). In quantum mechanics, we use the same definition, however the Poisson bracket is replaced by the commutator and $\{H,X_{a},Y_{b}\}$ are well-defined quantum mechanical operators, assumed to form an algebraically independent set.\par
%
%-------------------------------------------------------------------------------------------------------------------------------
%
The search for superintegrable Hamiltonians allowing second-order integrals of motion (i.e., quadratically superintegrable Hamiltonians) began in the mid 60ties with the work of Winternitz et al. in 2D Euclidian space \cite{Win1}. During the last ten years, the topic of superintegrability has become more popular and the classification of quadratically superintegrable systems has been extended in various directions, so that a large body of literature now exists \cite{Kal1,Kal2,Kal3,Kal4,Kal5,Kal6,Das1,Bal}. For a detailed list of references on quadratically superintegrable systems, we also refer the reader to the following review paper \cite{Win2}. Moreover, these quadratically superintegrable systems possess many interesting properties in classical and also in quantum mechanics. In addition, they are connected with various subjects in mathematical physics, such as exactly and quasi-exactly solvable systems and the well-known Hermite, Laguerre, and Jacobi classical orthogonal polynomials (COP) \cite{Erd}. More recently, the classification and study were pursued to systems with third-order integrals in 2D Euclidean space \cite{Win3,Gra,Mar1,Mar2,Tre,Pop}. If systems with third-order integrals of motion share many properties with the quadratically superintegrable ones, some of them such as the multiseparability or the fact that the classical and quantum systems coincide are lost. Some results are known about wavefunctions of these systems, however the connections with orthogonal polynomials remain to be understood.\par
%
%---------------------------------------------------------------------------------------------------------------------------------------
%
These works on superintegrability with third-order integrals pointed out that the direct approach, i.e., solving the corresponding overdetermined system of partial differential equations, would become more difficult to apply as the order increases. In order to circumvent these difficulties, many papers were devoted to constructing and classifying new superintegrable systems with higher-order integrals (i.e., the order of one of the integrals is greater than two) using other approaches, such as ladder operators \cite{Mar3}, recurrence relations \cite{Kal,Mil}, and supersymmetric quantum mechanics (SUSYQM) \cite{Mar4,Dem}. Many new families of superintegrable systems with higher-order integrals of motion were thus obtained and studied using these new methods.\par
%
%-----------------------------------------------------------------------------------------------------------------------------------
%
Independently of these works on superintegrability, many families of exceptional orthogonal polynomials (EOP) were obtained and used to construct new exactly solvable quantum potentials \cite{Gom1,Gom2,Gom3,Gom4,Fel,Sas1,Sas2,Sas3,Que1,Que2,Que3,Gra1,Gra2}. Very recently, it was recognized that these EOP could also be useful to construct new superintegrable systems with higher-order integrals \cite{Pos}. The Jacobi EOP and some related system were indeed used to generate new families of superintegrable systems. This approach needs further studies. The purpose of this paper is to obtain new superintegrable systems with higher-order integrals from Hermite and Laguerre EOP and to investigate the finite-dimensional unitary representations of their polynomial algebra, as well as the consequences on the degeneracies of energy levels.\par
%
%--------------------------------------------------------------------------------------------------------------------------------------
% 
Let us present the organization of this paper. In Sec. II, we recall some results concerning higher-order SUSYQM and polynomial Heisenberg algebras (PHA). In Sec. III, we review Hermite and Laguerre EOP and their related quantum systems. In Sec. IV, we recall some results on the construction of superintegrable systems from ladder operators. In Sec. V, from the 1D systems related to EOP, presented in Sec. III, and the method described in Sec. IV, we generate new families of superintegrable systems and we obtain their polynomial algebras and the finite-dimensional unitary representations of the latter. These results also enable us to understand the connection between EOP and two systems allowing second- and third-order integrals previously found by Gravel \cite{Gra}, as such systems are included in two of the new families constructed in the present paper.\par
%
%============================================================================
%
\section{Ladder operators, SUSYQM and polynomial Heisenberg algebras}

Le us consider a one-dimensional Hamiltonian $H^{(+)}$ (with $\hbar=1$ and $2m=1$),
\begin{equation}
  H^{(+)}=\frac{d^{2}}{dx^{2}}+V^{(+)}(x), \label{hamilp}
\end{equation}
possessing raising and lowering operators $a^{\dagger}$ and $a$ that are realized as $k$-th order differential operators
\begin{equation}
  a=\frac{d^{k}}{dx^{k}}+p_{k-1}(x)\frac{d^{k-1}}{dx^{k-1}}+\cdots+p_{1}(x)\frac{d}{dx}+p_{0}(x), 
  \label{ladderp}
\end{equation}
with the following commutation relations
\begin{equation}
  [H^{(+)},a]=-\lambda a,\quad [H^{(+)},a^{\dagger}]= \lambda a^{\dagger}, \label{phap1}
\end{equation}
\begin{equation}
  [a,a^{\dagger}]=P^{(+)}(H^{(+)}+\lambda)-P^{(+)}(H^{(+)}), \label{phap2}
\end{equation}
where $P^{(+)}$ is a $k$-th degree polynomial in $H^{(+)}$. This polynomial can be factorized in the following way
\begin{equation}
  P^{(+)}(H^{(+)})=\prod_{i=1}^{k}(H^{(+)}-\epsilon_{i}).  \label{functionpp}
\end{equation}
\par
%
%-----------------------------------------------------------------------------------------------------------------------------------
%
The PHA \cite{Fer,Car} generated by $\{H^{(+)},a,a^{\dagger}\}$ and defined by Eqs.~\eqref{phap1} and~\eqref{phap2} provides some information on the spectrum of $H^{(+)}$. The annihilation operator $a$ can allow at most $k$ zero modes (i.e., a state such that $a\psi=0$). As a consequence, by iteratively acting with the creation operator $a^{\dagger}$, at most $k$ infinite ladders can be generated. The creation operator can also allow zero modes (i.e., a state such $a^{\dagger}\psi=0$), in which case only finite sequences of levels (i.e., singlet, doublet, and more generally multiplet states) are obtained. The pattern of energy levels can become more complicated as the order of these ladder operators increases. In order to illustrate this statement, let us brieftly consider the $k=3$ case. There may exist three zero modes for the creation and annihilation operators, however, due to conflicting asymptotic properties, only three in total can be normalizable and thus acceptable wavefunctions. As a consequence, the pattern of levels consists in one, two or three infinite sequences of levels or an infinite sequence with a singlet state or doublet states \cite{Car,Mar1,And,Mat}. Some results are also known for $k=4$ \cite{Car,Mar5}. Let us notice the interesting fact that the cases of ladder operators of order three and four are related with the fourth and fifth Painlev\'e transcendents, respectively. However, ladder operators of order higher than four are an unexplored subject. In the next section, we will present systems related to EOP with ladder operators of order three, four, and six, respectively.\par
%
%------------------------------------------------------------------------------------------------------------------------------------
%
Let us now introduce a second 1D Hamiltonian $H^{(-)}$,
\begin{equation}
  H^{(-)}=-\frac{d^{2}}{dx^{2}}+V^{(-)}(x), \label{hamilm}
\end{equation}
related by higher-order SUSYQM with the initial Hamiltonian $H^{(+)}$ in the form
\begin{equation}
  f(H^{(+)})=A^{\dagger}A,\quad f(H^{(-)})=AA^{\dagger}, \label{functpm}
\end{equation}
\begin{equation}
  AH^{(+)}=H^{(-)}A,\quad A^{\dagger}H^{(-)}=H^{(+)}A. \label{interpm}
\end{equation}
The supercharges $A^{\dagger}$ and $A$ (of order $m$) are realized as $m$-th differential operators, and $A$ takes the form
\begin{equation}
  A=\frac{d^{m}}{dx^{m}}+q_{m-1}(x)\frac{d^{m-1}}{dx^{m-1}}+\cdots+q_{1}(x)\frac{d}{dx}+q_{0}(x). 
  \label{chargea}
\end{equation}
\par
%
%--------------------------------------------------------------------------------------------------------------------------------------
%
{}For first-order SUSYQM, Equation~\eqref{functpm} can be taken as
\begin{equation}
  f(H^{(+)})=H^{(+)}-E,\quad f(H^{(-)})=H^{(-)}-E, \label{functpm1}
\end{equation}
where $E$ is a factorization energy. The potentials are thus given by
\begin{equation}
  V^{(\pm)}=q_{0}^{2}(x)\mp q_{0}'(x)+E. \label{pot1}
\end{equation}
\par
%
%-------------------------------------------------------------------------------------------------------------------------------------
%
In the case of reducible second-order SUSYQM, the function $f$ can be taken as
\begin{equation}
  f(H^{(+)})=(H^{(+)})^{2}-\frac{c^{2}}{4},\quad f(H^{(-)})=(H^{(-)})^{2}-\frac{c^{2}}{4}. \label{functpm2}
\end{equation}
The potentials are thus given by
\begin{equation}
  V^{(\pm)}=\mp q_{1}'+\frac{q_{1}''}{2q_{1}}+\frac{q_{1}^{2}}{4}-\left(\frac{q_{1}'}{2q_{1}}\right)^{2}+
  \frac{c^{2}}{4q_{1}^{2}}, \label{pot2}
\end{equation}
where $c=E_{1}-E_{2}$ and $E_1$, $E_2$ are two factorization energies.\par
%
%------------------------------------------------------------------------------------------------------------------------------------
%
In the case of first-order SUSYQM (i.e., when $m=1$), the function $q_{0}$ is called the superpotential. The case of first-order supercharges was thoroughly studied \cite{Coo} and in particular shape-invariant systems were considered \cite{Gen}. The case of second-order supercharges was the object of many papers and, for instance, systems that are deformations of the harmonic oscillator were constructed \cite{Fern1,Fern2}. The applications of second-order SUSYQM to third-order ladder operators \cite{And,Mat,Mar1} and fourth-order ones \cite{Mar5,Car} were also studied. One can show that SUSYQM allows to relate wavefuntions of the Hamiltonians $H^{(+)}$ and $H^{(-)}$, their energy spectrum and also their ladder operators, as well as their corresponding PHA. Hence the Hamiltonian $H^{(-)}$ also admits a PHA of the form 
\begin{equation}
  [H^{(-)},b]=-\lambda b,\quad [H^{(-)},b^{\dagger}]=\lambda b^{\dagger}, \label{pham1}
\end{equation}
\begin{equation}
  [b,b^{\dagger}]=P^{(-)}(H^{(-)}+\lambda)-P^{(-)}(H^{(-)}), \label{pham2}
\end{equation}
with
\begin{equation}
  b=AaA^{\dagger},\quad b^{\dagger}=Aa^{\dagger}A^{\dagger}, \label{ladderm}
\end{equation}
\begin{equation}
  P^{(-)}(H^{(-)})=P^{(+)}(H^{(-)})f(H^{(-)}-\lambda)f(H^{(-)}). \label{pham3}
\end{equation}
This implies that SUSYQM can be used as a tool to construct systems with higher-order ladder operators.\par
%
%------------------------------------------------------------------------------------------------------------------------------------
%
In the next section, we will be interested in constructing 1D systems based on first- and second-order SUQYQM and  related to Hermite and Laguerre EOP.\par
%
%===========================================================================
%
\section{EOP and supersymmetric quantum mechanics}

The algebraic relations given by Eqs.~\eqref{functpm} and~\eqref{interpm} can be only formal because they do not take singularities nor boundaries into account. As a consequence, the isospectrality property can be lost as in the case of a regular Hamiltonian related to a singular superpartner Hamiltonian. It was shown that one can circumvent this problem by considering an appropriate nodeless seed solution (see, e.g., \cite{Coo,Fern1,Fern2}) to construct the supercharge operators given in Eq~\eqref{chargea} in order to obtain two regular superpartners (or two singular superpartners but with the same type and number of singularities), while SUSYQM relations are still valid. It was also shown in the case of first- and second-order SUSYQM that the exactly solvable systems obtained may be closely connected with EOP \cite{Gom1,Gom2,Gom3,Gom4,Fel,Sas1,Sas2,Sas3,Que1,Que2,Que3,Gra1,Gra2}.\par
%
%-----------------------------------------------------------------------------------------------------------------------------------
%
In the case of first-order SUSYQM, Equation~\eqref{functpm1} applies. One then uses a nodeless seed solution of the initial Schr\"odinger equation 
\begin{equation}
  \left(-\frac{d^{2}}{dx^{2}}+V^{(+)}(x)\right)\phi(x)=E\phi(x), \label{eqseed}
\end{equation}
with energy $E$ smaller than the ground-state energy $E_{0}^{(+)}$ of $H^{(+)}$, and therefore nonnormalizable. This choice differs from that made in``standard'' first-order SUSYQM, where the seed solution is the ground-state wavefunction. The superpotential is thus determined by
\begin{equation}
  q_{0}=-\frac{\phi'}{\phi}. \label{superpother}
\end{equation}
\par
%
%-------------------------------------------------------------------------------------------------------------------------------------
%
In the case of second-order SUSYQM, one considers two such seed solutions $\phi_{1},\phi_{2}$ of Eq.~\eqref{eqseed} with energy $E_{1}$ and $E_{2}$, respectively. The functions $q_{0}(x)$ and $q_{1}(x)$ in the supercharges $A$ and $A^{\dagger}$ given by~\eqref{chargea} (with $m=2$) take the form
\begin{equation}
  q_{1}=-\frac{\mathcal{W}'(\phi_{1},\phi_{2})}{\mathcal{W}(\phi_{1},\phi_{2})}, \quad  q_{0}=\frac{q_{1}'}{2}
  +\frac{q_{1}^{2}}{4}-\frac{q_{1}''}{2q_{1}}+\left(\frac{q_{1}'}{2q_{1}}\right)^{2}-\frac{c^{2}}{4q_{1}^{2}}, 
  \label{g0g1}
\end{equation}
where $\mathcal{W}(\phi_{1},\phi_{2})$ is the Wronskian of $\phi_{1}$ and $\phi_{2}$.\par
%
%+++++++++++++++++++++++++++++++++++++++++++++++++++++++++++++++++++++++++++
%
\subsection{Hermite EOP in first-order SUSYQM}

Let us consider as initial Hamiltonian $H^{(+)}$ (i.e.,~\eqref{hamilp}) the well-known harmonic oscillator, for which
\begin{equation}
  V^{(+)}=x^{2}  \label{oscip}
\end{equation}
is defined on the real line. The energy spectrum and the wavefunctions are given by
\begin{equation}
  \psi_{\nu}^{(+)}(x) \propto e^{-\frac{1}{2}x^{2}} H_{\nu}(x),\quad 
  E_{\nu}^{(+)}=2\nu+1,\quad \nu=0,1,2,\ldots, \label{enerosc}
\end{equation}
where $H_{\nu}(x)$ is the Hermite polynomial of degree $\nu$. This system possesses the following first-order ladder operators
\begin{equation}
  a=\frac{d}{dx}+x,\quad a^{\dagger}=-\frac{d}{dx}+x, \label{ladosc}
\end{equation}
which satisfy a PHA as given by Eqs.~\eqref{phap1} and~\eqref{phap2} with $\lambda=2$ and $P^{(+)}(H^{(+)})=H^{(+)}-1$.\par
%
%-------------------------------------------------------------------------------------------------------------------------------------
%
Let us consider first-order SUSYQM and a seed solution of the form \cite{Fel} 
\begin{equation}
  \phi_{m}(x)=(-i)^{m}H_{m}(ix)e^{\frac{1}{2}x^{2}}=\mathcal{H}_{m}(x)e^{\frac{1}{2}x^{2}}, \label{seedosc}
\end{equation}
with the corresponding energy $E_{m}=-(2m+1)$, where $\mathcal{H}_{m}(x)$ is a pseudo-Hermite polynomial \cite{Abr} of degre $m$ with all its coefficients positive. The seed solution is nodeless on the real line if we take $m=0,2,4,6,\ldots$. Furthermore, its inverse $\phi_{m}^{-1}(x)$ is  nonsingular and normalizable, so that it is an acceptable physical wavefunction of the superpartner potential.\par
%
%-----------------------------------------------------------------------------------------------------------------------------------
%
The superpotential $q_{0}$ can be written as 
\begin{equation}
  q_{0}=-x-\frac{\mathcal{H}_{m}'}{\mathcal{H}_{m}}. \label{superpotosc}
\end{equation}
On using identities satisfied by $\mathcal{H}_m(x)$, the  superpartner potential is obtained in the form
\begin{equation}
  V^{(-)}(x)=x^{2}-2\left[\frac{\mathcal{H}_{m}''}{\mathcal{H}_{m}}-\left(\frac{\mathcal{H}_{m}'}
  {\mathcal{H}_{m}}\right)^{2}+1\right], \label{partnerosc}
\end{equation}
with the energy spectrum
\begin{equation}
  E_{\nu}^{(-)}=2\nu+1,\quad \nu=-m-1,0,1,2,\ldots, \label{enerpartosc}
\end{equation}
and the corresponding wavefunctions 
\begin{equation}
  \psi_{\nu}^{(-)}(x) \propto \left\{
    \begin{array}{ll}
    \phi_{m}^{-1}(x) & {\rm if\ } \nu=-m-1, \\[0.2cm] 
    A \psi_{\nu}^{(+)}(x) & {\rm if\ } \nu=0, 1, 2,\ldots, 
    \end{array}
  \right.  \label{wavepartosc2}
\end{equation}
which can be rewritten as
\begin{equation}
  \psi_{\nu}^{(-)}(x) \propto \frac{e^{-\frac{1}{2}x^{2}}}{\mathcal{H}_{m}}y_n(x), \quad \nu=-m-1, 0, 1, 2,\ldots  
\end{equation}
The latter involve the Hermite EOP $y_{n}(x)$, which are $n$th-degree polynomials (with $n=m+\nu +1$) forming an orthogonal and complete set with respect to the positive-definite measure $e^{-x^2}\bigl(\mathcal{H}_m(x)\bigr)^{-2} dx$. Such polynomials are defined by
\begin{equation}
  y_n(x) = \left\{
    \begin{array}{ll}
    1 & {\rm if\ } \nu=-m-1, \\[0.2cm]
    - \mathcal{H}_m H_{\nu+1} - 2m \mathcal{H}_{m-1} H_{\nu} & {\rm if\ } \nu=0, 1, 2,\ldots,  
    \end{array}
  \right.  \label{EOPhermite}
\end{equation}
and satisfy the second-order differential equation
\begin{equation}
  \left[\frac{d^2}{dx^2} - 2 \left(x + \frac{\mathcal{H}'_m}{\mathcal{H}_m}\right) \frac{d}{dx} + 2n\right] 
  y_n(x) = 0.  \label{diff-eqn}
\end{equation}
\par
%
%-----------------------------------------------------------------------------------------------------------------------------------
%
The system obtained using SUSYQM and given by Eq.~\eqref{partnerosc} has ladder operators of third order, which are constructed using the ladder operators \eqref{ladosc} of the initial Hamiltonian and the supercharges, as shown in Eq.~\eqref{ladderm}. These ladder operators also satisfy a PHA given by Eq.~\eqref{pham3} with $P^{(-)}(H^{(-)})=(H^{(-)}-1)(H^{(-)}-E)(H^{(-)}-E-2)$ and $\lambda=2$, where $E$ is the factorization energy $E_m$.\par
%
%+++++++++++++++++++++++++++++++++++++++++++++++++++++++++++++++++++++++++++
%
\subsection{Laguerre EOP in first-order SUSYQM}

Let us now consider the case of the radial oscillator, for which
\begin{equation}
  V_{l}(x)=\frac{1}{4}x^{2}+\frac{l(l+1)}{x^{2}} \label{singoscpot}
\end{equation}
is defined on the positive half-line. The wavefunctions (with $z=\frac{1}{2}x^{2}$, $\alpha=l+\frac{1}{2}$, $\nu=0$, 1, 2, \ldots) are given by
\begin{equation}
  \psi_{\nu l}(x) \propto x^{l+1}e^{-\frac{1}{4}x^{2}}L_{\nu}^{(l+\frac{1}{2})}\left(\frac{1}{2}x^{2}\right)
  \propto \eta_{l}(z)L_{\nu}^{(\alpha)}(z), \quad \eta_l(z) = z^{\frac{1}{4}(2\alpha+1)} e^{-\frac{1}{2}z}, 
  \label{singoscwave}
\end{equation}
with the corresponding energy spectrum
\begin{equation}
  E_{\nu l}=2\nu +l+\frac{3}{2}. \label{singoscenerg}
\end{equation}
\par
%
%---------------------------------------------------------------------------------------------------------------------------------------
%
In this case we have three possible nonnormalizable  seed solutions $\phi_{lm}(x)$ \cite{Que2,Que3,Gra1}, which we present in Table I together with their corresponding energy $E_{lm}$. 

\begin{table}[h]
\begin{center}
\begin{tabular}{|l|l|l|l|}
\hline
  &   $\phi_{lm}$ &  $\chi_{l}$  & $E_{lm}$  \\
\hline
  Case I  & $\chi_{l}^{\rm I}(z)L_{m}^{(\alpha)}(-z)$  & $z^{\frac{1}{4}(2\alpha +1)}e^{\frac{1}{2}z}$  &  
        $-(\alpha +2m+1)$  \\
\hline
  Case II  &  $\chi_{l}^{\rm II}(z)L_{m}^{(-\alpha)}(z)$ & $z^{-\frac{1}{4}(2\alpha -1)}e^{-\frac{1}{2}z}$ &  
        $-(\alpha-2m-1)$ \\  
\hline
  Case III  & $\chi_{l}^{\rm III}(z)L_{m}^{(-\alpha)}(-z)$  & $z^{-\frac{1}{4}(2\alpha -1)}e^{\frac{1}{2}z}$ &
        $\alpha-2m-1$  \\  
\hline
\end{tabular}
\caption{\label{tab:5/tc} Seed solutions: Cases I,II, and III}
\end{center}
\vspace{-0.6cm}
\end{table}

%
%-----------------------------------------------------------------------------------------------------------------------------------------
%
They can be used to construct a superpartner $V^{(-)}(x)$ in first-order SUSYQM. To be able to write the latter in a unified manner as 
\begin{equation}
  V^{(-)}(x)=V_{l}(x) + V_{l,\rm rat}(x)+C, \label{singoscisuperpart1}
\end{equation}
\begin{equation}
  V_{l,\rm rat}(x)=-2\left\{\frac{\dot{g}_m^{(\alpha)}}{g_{m}^{(\alpha)}}+2z \left[ 
  \frac{\ddot{g}_{m}^{(\alpha)}}{g_{m}^{(\alpha)}}-\left(\frac{\dot{g}_{m}^{(\alpha)}}{g_{m}^{(\alpha)}}
  \right)^{2}\right]\right\}, \label{singoscisuperpart2}
\end{equation}
where a dot denotes a derivative with respect to $z$, it is convenient to start from a potential \eqref{singoscpot} with a different value $l'$ of the angular momentum, which depends on the case considered,
\begin{equation}
  V^{(+)}(x) = V_{l'}(x).  \label{singoscpot1}
\end{equation}
Observe that a similar change $l \to l'$ has to be done in the wavefunctions \eqref{singoscwave} and the energy spectrum \eqref{singoscenerg}, so that $\psi^{(+)}_{\nu l}(x) = \psi_{\nu l'}(x)$ and $E^{(+)}_{\nu l} = E_{\nu l'}$.\par
%
%-----------------------------------------------------------------------------------------------------------------------------------
%
In Table II, we present for each case the values of $l'$, of the constant $C$, and of $m$, as well as the polynomial $g_{m}^{(\alpha)}(z)$ occurring in Eq.~\eqref{singoscisuperpart2}, the constraints on $\alpha$ and $m$, and the resulting energy spectrum $E^{(-)}_{\nu l}$. 

\begin{table}[h]
\begin{center}
\begin{tabular}{|l|l|l|l|l|l|l|}
\hline
  & $l'$ &  $g_{m}^{(\alpha)}$   &  $C$ &  $m$  &  Constraints & $E^{(-)}_{\nu l}$ \\
\hline
  Case I & $l-1$  & $L_{m}^{(\alpha-1)}(-z)$ & $-1$  & $1,2,3,\ldots$  &     & $2\nu+\alpha$, 
       $\nu=0,1,2,\ldots$ \\
\hline
  Case II & $l+1$  & $L_{m}^{(-\alpha-1)}(z)$  &  $1$ & $1,2,3,\ldots$ & $\alpha > m-1$ & $2\nu+\alpha+2$, 
       $\nu=0,1,2,\ldots$ \\  
\hline
  Case III & $l+1$ & $L_{m}^{(-\alpha-1)}(-z)$  & $-1$  & $2,4,6,\ldots$ & $\alpha > m-1$  & $2\nu+\alpha+2$, 
       $\nu=-m-1,0,1,2,\ldots$  \\  
\hline
\end{tabular}
\caption{\label{tab:5/tc} Superpartner: Cases I,II, and III}
\end{center}
\vspace{-0.6cm}
\end{table}

%
%------------------------------------------------------------------------------------------------------------------------------------
%
The bound-state wavefunctions $\psi_{\nu l}^{(-)}(x)$ of $H^{(-)}$ are given by
\begin{equation}
  \psi_{\nu l}^{(-)}(x) \propto \frac{\eta_{l}(z)}{g_{m}^{(\alpha)}(z)}y_{n}^{(\alpha)}(z), 
  \label{singoscipartwave1}
\end{equation}
where in Cases I and II, $n=m+\nu$ with $\nu=0,1,2,\ldots$, while in Case III, $n=m+\nu+1$ with $\nu=-m-1,0,1,2,\ldots$. They can be obtained by acting with $A$ on those of $H^{(+)}$, except for case III and $\nu=-m-1$, where $\psi^{(-)}_{-m-1,l}(x) \propto \left(\phi^{\rm III}_{lm}(x)\right)^{-1}$ and therefore $y^{(\alpha)}_0(z)=1$. In all cases, the $n$-th degree polynomials $y_{n}^{(\alpha)}(z)$ satisfy the following second-order differential equation
\begin{equation}
  \left[z\frac{d^{2}}{dz^{2}}+\left(\alpha+1-z-2z\frac{\dot{g}_{m}^{(\alpha)}}{g_{m}^{(\alpha)}}\right)
  \frac{d}{dz}+(z-\alpha)\frac{\dot{g}_{m}^{(\alpha)}}{g_{m}^{(\alpha)}}
  +z\frac{\ddot{g}_{m}^{(\alpha)}}{g_{m}^{(\alpha)}}\right]y_{n}^{(\alpha)}(z)=(m-n)y_{n}^{(\alpha)}(z) .
\end{equation} \label{singoscipartwave2}
Moreover, they form an orthogonal and complete set with respect of the positive-definite measure $z^{\alpha}e^{-z}\bigl(g_{m}^{(\alpha)}(z)\bigr)^{-2}dz$. As a consequence, there exist three families of Laguerre EOP, which are denoted as $L_{\alpha,m,n}^{\rm I}(z)$, $L_{\alpha,m,n}^{\rm II}(z)$, and $L_{\alpha,m,n}^{\rm III}(z)$, respectively.\par
%
%----------------------------------------------------------------------------------------------------------------------------------
%
The radial oscillator \eqref{singoscpot1} has second-order ladder operators 
\begin{equation}
\begin{split}
  a=\frac{1}{4}\left(2\frac{d^{2}}{dx^{2}}+2x\frac{d}{dx}+\frac{1}{2}x^{2}-\frac{2l'(l'+1)}{x^{2}}+1\right), \\ 
  a^{\dagger}=\frac{1}{4}\left(2\frac{d^{2}}{dx^{2}}-2x\frac{d}{dx}+\frac{1}{2}x^{2}-\frac{2l'(l'+1)}{x^{2}}-1
  \right),
\end{split} \label{singosclad}
\end{equation}
which satisfy a PHA as given by~\eqref{phap1} and~\eqref{phap2} with $\lambda=2$ and $P^{(+)}(H^{(+)})=\frac{1}{16}( 2 H^{(+)}-3 - 2l')(2H^{(+)}-1+2l')$. Thus the three possible superpartners have ladder operators of order four as shown in~\eqref{ladderm}. They possess a PHA, as given by~\eqref{pham1},~\eqref{pham2}, and~\eqref{pham3} with $\lambda=2$ and $P^{(-)}(H^{(-)})=\frac{1}{16}( 2 H^{(-)}-3 - 2l')(2H^{(-)}-1+2l')(H^{(-)}-2-E)(H^{(-)}-E)$, where $E$ is the factorization energy $E_{lm}$ given in Table I with $\alpha$ replaced by $\alpha-1$ in Case I and by $\alpha+1$ in Case II or III.\par
%
%+++++++++++++++++++++++++++++++++++++++++++++++++++++++++++++++++++++++++
%
\subsection{Laguerre EOP in second-order SUSYQM}

In the case of systems constructed using reducible second-order SUSYQM associated with EOP, we need to consider two nodeless seed solutions $\phi_{1}$ and $\phi_{2}$. For the radial oscillator, we have three types of seed solutions and thus in the case of second-order SUSYQM we get six different possibilities.\par
%
%--------------------------------------------------------------------------------------------------------------------------------------
%
Let us restrict ourselves to one of the six cases, i.e., when the first seed solution $\phi_{1}$ is taken of type I and the second $\phi_{2}$ of type II, as presented in Tables I and II. For this choice, we have
\begin{equation}
  V^{(+)}(x)=V_{l}(x)-\frac{1}{2}(E_{1}+E_{2}),\quad \phi_{1}(x)=\phi_{l m_{1}}^{\rm I}(x),\quad 
  \phi_2(x) = \phi_{l m_{2}}^{\rm II}(x), \label{singosciorder2}
\end{equation}
where the energies associated with the two seed solutions are
\begin{equation}
  E_{1}=-\left(l+2m_{1}+\frac{3}{2}\right),\quad E_{2}=-\left(l-2m_{2}-\frac{1}{2}\right),\quad 
  m_{2}<l+\frac{1}{2}. \label{E1E2}
\end{equation}
The supercharges for second-order SUSYQM, given in Eqs.~\eqref{chargea} and \eqref{g0g1}, can be constructed from the Wronskian
\begin{equation}
  \mathcal{W}(\phi_{1},\phi_{2})= \frac{2}{x}g^{(\alpha)}_{\mu}(z)\chi_{l}^{\rm I}(z)\chi_{l}^{\rm II}(z), 
  \label{functionW}
\end{equation}
\begin{equation}
  g^{(\alpha)}_{\mu}=z \mathcal{\tilde{W}}\bigl(L_{m_{1}}^{(\alpha)}(-z),L_{m_{2}}^{(-\alpha)}(z)\bigr)
  -(z+\alpha)L_{m_{1}}^{(\alpha)}(-z)L_{m_{2}}^{(-\alpha)}(z), \label{functiong}
\end{equation}
where $g^{(\alpha)}_{\mu}$ is a $\mu$th-degree polynomial with $\mu = m_1+m_2+1$ and $\tilde{\cal W}\bigl(f(z), g(z)\bigr)$ denotes the Wronskian of two $z$-dependent functions. The superpartner takes the form
\begin{equation}
  V^{(-)}(x)=V_{l}(x)-2\left\{\frac{\dot{g}^{(\alpha)}_{\mu}}{g^{(\alpha)}_{\mu}}
  +2z\left[\frac{\ddot{g}^{(\alpha)}_{\mu}}{g^{(\alpha)}_{\mu}}
  -\left(\frac{\dot{g}^{(\alpha)}_{\mu}}{g^{(\alpha)}_{\mu}}\right)^{2}\right]\right\}-\frac{1}{2}(E_{1}+E_{2}) 
  \label{singoscorder2part}
\end{equation}
and its energy spectrum, which can be calculated from SUSYQM, is the following
\begin{equation}
  E_{\nu l}^{(-)}=E_{\nu l}^{(+)}=2\nu +2l +m_{1}-m_{2}+2,\quad \nu=0,1,2, \ldots. 
  \label{singoscorder2partener}
\end{equation}
\par
%
%---------------------------------------------------------------------------------------------------------------------------------
%
The bound-state wavefunctions of $H^{(-)}$ can be obtained by acting with $A$ on those of $H^{(+)}$, $\psi^{(+)}_{\nu l}(x) = \psi_{\nu l}(x)$, and are given by
\begin{equation}
  \psi_{\nu l}^{(-)}(x) \propto \frac{\eta_{l}(z)}{g_{\mu}^{(\alpha)}(z)}y_{n}^{(\alpha)}(z),\quad n=\mu+\nu,
  \quad \nu=0,1,2,\ldots, \label{singoscipartwave1bis}
\end{equation}
where the $n$-th degree polynomial $y_{n}^{(\alpha)}(z)$ satisfies the second-order differential equation 
\begin{equation}
  \left[z \frac{d^2}{dz^2} + \left(\alpha + 1 - z - 2z \frac{\dot{g}_{\mu}^{(\alpha)}}{g_{\mu}^{(\alpha)}}
  \right) \frac{d}{dz} + (z - \alpha) \frac{\dot{g}_{\mu}^{(\alpha)}}{g_{\mu}^{(\alpha)}} + z 
  \frac{\ddot{g}_{\mu}^{(\alpha)}}{g_{\mu}^{(\alpha)}}\right] y_n^{(\alpha)}(z) = (\mu - n)
  y_n^{(\alpha)}(z), \label{singoscorderwave}
\end{equation}
which is similar to the corresponding result \eqref{singoscipartwave2} obtained in first-order SUSYQM. It can be shown that the polynomials $y_{n}^{(\alpha)}(z)$ form an orthogonal and complete set with respect of the positive-definite measure $z^{\alpha}e^{-z}\bigl(g_{\mu}^{(\alpha)}\bigr)^{-2}dz$. These EOP are denoted by
$L_{\alpha,m_{1},m_{2},n}^{\rm I,II}(z)$.\par
%
%--------------------------------------------------------------------------------------------------------------------------------------
%
{}For convenience, we remove the additive constant $\frac{1}{2}(E_{1}+E_{2})$ in the potentials $V^{(+)}(x)$ and $V^{(-)}(x)$. This translation will only affect the energies \eqref{singoscorder2partener}, but not the superintegrability property. Taking this modification into account, we find the following factorization relations $f(H^{(+)})=(H^{(+)}-E_{1})(H^{(+)}-E_{2})$ and $f(H^{(-)})=(H^{(-)}-E_{1})(H^{(-)}-E_{2})$ for the functions defined in Eq.~\eqref{functpm}. The superpartner $H^{(-)}$ has ladder operators that are realized as differential operators of order six and obtained from Eq.~\eqref{ladderm}. The PHA of $H^{(-)}$ is given by~\eqref{pham1},~\eqref{pham2}, and~\eqref{pham3} with $\lambda=2$ and $P^{(-)}(H^{(-)})=\frac{1}{16}( 2 H^{(-)}-3 - 2l)(2H^{(-)}-1+2l)(H^{(-)}-E_{1}-2)(H^{(-)}-E_{2}-2)(H^{(-)}-E_{1})(H^{(-)}-E_{2})$.\par
%
%===========================================================================
%
\section{Superintegrability and ladder operators}

The direct approach to obtain superintegrable systems with higher-order integrals of motion leads to overdetermined systems of partial differential equations. The corresponding compatibility equations take the form of nonlinear differential equations, which may be a challenging problem to solve. A way to generate new superintegrable systems is to use 1D systems allowing ladder operators (or recurrence relations) and to construct multidimensional systems with integrals of motion that take the form of products of ladder operators. Let us mention that the classification of systems with ladder operators can also be challenging, however, as seen in Sec.~II, SUSYQM provides a way to obtain systems with higher-order ladder operators from specific superpartners.
\par
%
%-------------------------------------------------------------------------------------------------------------------------------------
%
Let us consider the following two-dimensional Hamiltonian allowing separation of variables in Cartesian coordinates:
\begin{equation}
  H=H_{x}+H_{y}=-\frac{d^{2}}{dx^{2}}-\frac{d^{2}}{dy^{2}}+V_{x}(x)+V_{y}(y). \label{hamil2d}
\end{equation}
We impose the existence of ladder operators of form~\eqref{ladderp} in both axes that satisfy the PHA
\begin{subequations}
\begin{equation}
  [H_{x},a_{x}^{\dagger}]=\lambda_{x}a_{x}^{\dagger},\quad [H_{x},a_{x}]=-\lambda_{x}a_{x}, \label{phax1}
\end{equation}
\begin{equation}
  a_{x}a_{x}^{\dagger}=Q(H_{x}+\lambda_{x}),\quad a_{x}^{\dagger}a_{x}=Q(H_{x}), \label{phax2}
\end{equation}
\end{subequations}
\begin{subequations}
\begin{equation}
  [H_{y},a_{y}^{\dagger}]=\lambda_{y}a_{y}^{\dagger},\quad [H_{y},a_{y}]=-\lambda_{y}a_{y}, \label{phay1}
\end{equation}
\begin{equation}
  a_{y}a_{y}^{\dagger}=S(H_{y}+\lambda_{y}),\quad a_{y}^{\dagger}a_{y}=S(H_{y}), \label{phay2}
\end{equation}
\end{subequations}
where $\lambda_{x}$ and $\lambda_{y}$ are constants while $Q(x)$ and $S(y)$ are polynomials. From these operators we get the following integrals of motion (of order 2, $k_{1}n_{1}+k_{2}n_{2}$, and $k_{1}n_{1}+k_{2}n_{2}$) with $n_{1}\lambda_{x}=n_{2}\lambda_{y}=\lambda$, $n_{1}$,$n_{2}$ $\in \mathbb{Z}^{*}$,
\begin{equation}
  K=\frac{1}{2\lambda}(H_{x}-H_{y}),\quad I_{-}=a_{x}^{n_{1}}a_{y}^{\dagger n_{2}},\quad I_{+}=a_{x}^{\dagger n_{1}}a_{y}^{n_{2}}. \label{KI}
\end{equation}
The system possesses three algebraically independent integrals of motion and is thus maximally superintegrable. We can also consider the integrals $I_{1}=I_{-}-I_{+}$ and $I_{2}=I_{-}+I_{+}$. Let us point out an important aspect of this method. From the equations above, we can see that even with ladder operators of lower order, the method allows to generate integrals of motion of an arbitrary order in a nice factorized form that would be difficult to obtain in a direct approach.\par
%
%-----------------------------------------------------------------------------------------------------------------------------------
%
These integrals of motion generate the polynomial algebra of the system,
\begin{equation}
  [K,I_{\pm}]=\pm I_{\pm}, \quad [I_{-},I_{+}]=F_{n_{1},n_{2}}(K+1,H)-F_{n_{1},n_{2}}(K,H), \label{KIcomm}
\end{equation}
\begin{equation}
  F_{n_{1},n_{2}}(K,H)=\prod_{i=1}^{n_{1}}Q\left(\frac{H}{2}+\lambda K-(n_{1}-i)\lambda_{x}\right)
  \prod_{j=1}^{n_{2}}S\left(\frac{H}{2}-\lambda K+j\lambda_{y}\right), \label{functionF}
\end{equation}
which is of order $k_{1}n_{1}+k_{2}n_{2}-1$. Such a polynomial algebra is a deformed $u(2)$ algebra and its finite-dimensional representation modules can be found by realizing it as a generalized deformed oscillator algebra $\{b^t, b, N\}$. The operators $b^t=I_{+}$, $b=I_{-}$, $N=K-u$ and $\Phi(H,u,N)=F_{n_{1},n_{2}}(K,H)$ indeed satisfy the defining relations of such an algebra \cite{Das2},
\begin{equation}
  [N,b^{t}]=b^{t} ,\quad [N,b]=-b,\quad b^{t}b=\Phi(H,u,N),\quad bb^{t}=\Phi(H,u,N+1), \label{bbdncommu}
\end{equation}
where $u$ is some constant and $\Phi(H,u,N)$ is called ``structure function''. If the latter satisfies the properties
\begin{equation}
  \Phi(E,u,0)=0 ,\quad \Phi(E,u,p+1)=0, \quad \Phi(E,u,n) >0 \quad  n=1,2,\ldots,p, \label{constraints}
\end{equation}
then the deformed oscillator algebra has an energy-dependent Fock space of dimension $p+1$ with a Fock basis $|E,n\rangle$, $n=0$, 1,~\ldots, $p$, fulfilling
\begin{subequations}
\begin{equation}
  H|E,n\rangle=E|E,n\rangle,\quad N|E,n\rangle=n|E,n\rangle, \quad b|E,0\rangle=0, \quad b^t|E,p\rangle = 0, 
  \label{Fock1}
\end{equation}
\begin{equation}
  b^{t}|E,n\rangle=\sqrt{\Phi(E,u,n+1)}|E,n+1\rangle,\quad b|n\rangle=\sqrt{\Phi(E,u,n)}|E,n-1\rangle. \label{Fock2}
\end{equation}
\end{subequations}
These relations can be used to obtain the finite-dimensional unitary representations of the polynomial algebra \eqref{KIcomm}, \eqref{functionF}, and the corresponding degenerate energy spectrum of the system.
In the next section, we will restrict the construction of superintegrable systems to the case $\lambda_{x}=\lambda_{y}$, wherein each of the products contained in \eqref{functionF} reduces to only one term. This will allow to identify more clearly new degeneracies not obtained in the finite-dimensional unitary representations. Note, however, that all the families of superintegrable systems considered in this paper could be extended in a straightforward manner to more general cases where $\lambda_x \ne \lambda_y$.\par
%
%===========================================================================
% 
\section{New families of 2D superintegrable systems and polynomial algebras from EOP}

\subsection{Superintegrable systems from Hermite EOP}

\textbf{Case 1}

Let us consider the two-dimensional superintegrable system given by Eq.~\eqref{hamil2d} with respectively in the $x$-axis the superpartner of the harmonic oscillator related to Hermite EOP presented in Sec.~IIIA and in the $y$-axis the harmonic oscillator itself,
\begin{equation}
  H_{x}=-\frac{d^{2}}{dx^{2}} +  x^{2}-2\left[\frac{\mathcal{H}_{m}''}{\mathcal{H}_{m}}
  -\left(\frac{\mathcal{H}_{m}'}{\mathcal{H}_{m}}\right)^{2}+1\right], \quad m {\ \rm even}, \label{sys1x}
\end{equation}
\begin{equation}
  H_{y}=-\frac{d^{2}}{dy^{2}} +  y^{2}. \label{sys1y}
\end{equation}
This 2D system includes one of the Gravel's systems \cite{Gra} (Potential 1 in Ref.~\cite{Mar2}) for $m=2$. It has
integrals of motion given by~\eqref{KI} and the energy spectrum corresponding to physical states is given by
\begin{equation}
  E=E_{x}+E_{y}=2(\nu_{x}+\nu_{y}+1), \quad \nu_{x}=-m-1,0,1,2,\ldots,\quad \nu_{y}=0,1,2,\dots. \label{energ1}
\end{equation}
\par
%
%___________________________________________________________________________
%
From Sec.~IIIA, we know that in both axes the ladder operators satisfy a PHA. Equations~\eqref{phax1}--\eqref{phay2} therefore apply with $\lambda_{x}=\lambda_{y}=2$ and 
\begin{equation}
  Q(H_{x})=(H_{x}-1)(H_{x}+2m-1)(H_{x}+2m+1),\quad S(H_{y})=H_{y}-1. \label{QS1}
\end{equation}
The structure function $\Phi(E,u,x)$ is obtained from Eqs.~\eqref{functionF} and~\eqref{QS1} as
\begin{equation}
\begin{split}
  \Phi(E,u,x)&=\left(\frac{E}{2}+2x+2u-1\right)\left(\frac{E}{2}+2x+2u+2m-1\right) \\
  &\quad \times\left(\frac{E}{2}+2x+2u+2m+1\right) \left(\frac{E}{2}-2x-2u+1\right). \label{struc1}
\end{split}
\end{equation}
From this structure function and the first constraint of Eq.~\eqref{constraints}, we obtain four solutions for the parameter $u$,
\begin{equation}
  u_{1}=-\frac{E}{4}+\frac{1}{2},\quad u_{2}=-\frac{E}{4}-m+\frac{1}{2},\quad u_{3}=-\frac{E}{4}-m-\frac{1}{2},  
  \quad u_{4}=\frac{E}{4}+\frac{1}{2}. \label{paramu1}
\end{equation}
The finite-dimensional unitary representations are calculated from the two other constraints of Eq.~\eqref{constraints} and are presented in Table III.

\begin{table}[h]
\begin{center}
\begin{tabular}{|l|l|l|l|l|l|}
\hline
  & $u$   &  $p$ &  Energy $E$ &  Structure function $\Phi$  &  Physical states   \\
\hline
  1 & $u_{1}$  & $\mathbb{N}$  &  $2(p+1)$  & $16x(p+1-x)(x+m)(x+1+m)$ & $\nu_{x}=0, 1, 2,\ldots$, 
      $\nu_{y}=0,1,2,\ldots$  \\
\hline
  2 & $u_{3}$   & 0  &  $2(p-m)$ & $16x(p+1-x)(x-1-m)(x-1)$ &   $\nu_{x}=-m-1$, $\nu_{y}=0$  \\  
\hline
\end{tabular}
\caption{\label{tab:5/tc} Finite-dimensional unitary representations for the superintegrable system given by Eqs.~\eqref{sys1x} and~\eqref{sys1y} and related to Hermite EOP.}
\end{center}
\vspace{-0.6cm}
\end{table}

%
%--------------------------------------------------------------------------------------------------------------------------------------
%
As shown by Table III, in terms of physical states, the first solution corresponds to all excited states of Hamiltonian $H_{x}$ combined with all states of Hamiltonian $H_{y}$, while the second solution is associated with the ground state of $H_{x}$ combined with that of $H_{y}$.\par
%
%xxxxxxxxxxxxxxxxxxxxxxxxxxxxxxxxxxxxxxxxxxxxxxxxxxxxxxxxxxxxxxxxxxxxxxxxxxxxxxxxxxxxxxxxxxxxxxxx
%
\textbf{Case 2}

Let us now consider the following 2D superintegrable system with both potentials related to Hermite EOP, 
\begin{equation}
  H_{x}=-\frac{d^{2}}{dx^{2}} +  x^{2}-2\left[\frac{\mathcal{H}_{m_{1}}''}{\mathcal{H}_{m_{1}}}
  -\left(\frac{\mathcal{H}_{m_{1}}'}{\mathcal{H}_{m_{1}}}\right)^{2}+1\right], \label{sys2x}
\end{equation}
\begin{equation}
  H_{y}=-\frac{d^{2}}{dy^{2}} +  y^{2}-2\left[\frac{\mathcal{H}_{m_{2}}''}{\mathcal{H}_{m_{2}}}
  -\left(\frac{\mathcal{H}_{m_{2}}'}{\mathcal{H}_{m_{2}}}\right)^{2}+1\right], \label{sys2y}
\end{equation}
where $m_1$ and $m_2$ are even and we may assume $m_1 \ge m_2$. This is a generalization of Case 1, which also includes another system obtained by Gravel \cite{Gra} (Potential 6 of Ref.~\cite{Mar2}) for $m_{1}=m_{2}=2$.
The integrals of motion are given by Eq.~\eqref{KI} again and the energy spectrum of physical states is
\begin{equation}
  E=E_{x}+E_{y}=2(\nu_{x}+\nu_{y}+1), \quad \nu_{x}=-m_1-1,0,1,\ldots,\quad \nu_{y}=-m_2-1,0,1,\ldots. 
  \label{ener2}
\end{equation}
\par
%
%------------------------------------------------------------------------------------------------------------------------------------
%
{}From Sec.~IIIA, we also know that in both axes the ladder operators satisfy a PHA. In Eqs.~\eqref{phax1}--\eqref{phay2}, we have $\lambda_{x}=\lambda_{y}=2$ and 
\begin{equation}
\begin{split}
  Q(H_{x})&=(H_{x}-1)(H_{x}+2m_{1}-1)(H_{x}+2m_{1}+1), \\ 
  S(H_{y})&=(H_{y}-1)(H_{y}+2m_{2}-1)(H_{y}+2m_{2}+1).
\end{split}  \label{QS2}
\end{equation}
The structure function calculated from~\eqref{functionF} and~\eqref{QS2} is
\begin{equation}
\begin{split}
  \Phi(E,u,x)&=\left(\frac{E}{2}+2x+2u-1\right)\left(\frac{E}{2}+2x+2u+2m_{1}-1\right) \\
  & \quad \times\left(\frac{E}{2}+2x+2u+2m_{1}+1\right) \left(\frac{E}{2}-2x-2u+1\right) \\
  & \quad \times\left(\frac{E}{2}-2x-2u+2m_{2}+1\right)\left(\frac{E}{2}-2x-2u+2m_{2}+3\right).
\end{split}  \label{functionF2}
\end{equation}
%
%------------------------------------------------------------------------------------------------------------------------------------
%
{}From the first constraint of Eq.~\eqref{constraints}, we find six solutions for the parameter $u$,
\begin{equation}
\begin{split}
  u_{1}&=-\frac{E}{4}+\frac{1}{2},\quad u_{2}=-\frac{E}{4}-m_{1}+\frac{1}{2},\quad u_{3}=-\frac{E}{4}-m_{1}-
     \frac{1}{2}, \\
  u_{4}&=\frac{E}{4}+\frac{1}{2},\quad u_{5}=\frac{E}{4}+m_{2}+\frac{1}{2},\quad u_{6}=\frac{E}{4}+m_{2}+
     \frac{3}{2}.
\end{split} \label{paramu2}
\end{equation}
The two other constraints of Eq.~\eqref{constraints} allow to obtain the finite-dimensional unitary representations, which are given in Table IV.

\begin{table}[h]
\begin{center}
\begin{tabular}{|l|l|l|l|l|l|}
\hline
  & $u$   &  $p$ &  Energy $E$  &  Structure function $\Phi$  &  Physical states   \\
\hline
  1 & $u_{1}$  & $\mathbb{N}$  &  $2(p+1)$  & $64x(p+1-x)(x+m_{1})(x+1+m_{1})$ & $\nu_{x}=0,1,2,\ldots$, 
       \\
   &  &  &   & ${}\times (p+1+m_{2}-x)(p+2+m_{2}-x)$ & $\nu_{y}=0,1,2,\ldots$  \\  
\hline
  2 & $u_{1}$   & 0  &  $2(p-m_{2})$ & $64x(p+1-x)(p-x)(x+m_{1})$ &   $\nu_{x}=0$,   \\ 
  &    &   &   & ${}\times (x+1+m_{1})(p-m_{2}-x)$ & $\nu_{y}=-m_{2}-1$   \\   
\hline
  3 & $u_{3}$   & 0  &  $2(p-1-m_{1}-m_{2})$ & $64x(p+1-x)(p-x)(x-1-m_{1})$ &   $\nu_{x}=-m_{1}-1$,  \\    
  &    &   &   & ${}\times (x-1)(p-m_{2}-x)$ &  $\nu_{y}=-m_{2}-1$  \\
\hline
  4 & $u_{3}$   & 0  &  $2(p-m_{1})$ & $64x(x-1)(x-1-m_{1})(p+1-x)$ &   $\nu_{x}=-m_{1}-1$,  \\ 
   &   &   &   & ${}\times (p+1+m_{2}-x)(p+2+m_{2}-x)$ & $\nu_{y}=0$    \\   
\hline
  5 & $u_{5}$   & 0  &  $-2(p+1+m_{1}+m_{2})$ & $64x(p+1-x)(x-1)(x+m_{2})$ &   $\nu_{x}=-m_{1}-1$, \\  
  &   &   &  & ${}\times (p-x)(p+1+m_{1}-x)$ & $\nu_{y}=-m_{2}-1$   \\
\hline
\end{tabular}
\caption{\label{tab:5/tc} Finite-dimensional unitary representations for the superintegrable system given by Eqs.~\eqref{sys2x} and~\eqref{sys2y} and related to Hermite EOP.}
\end{center}
\vspace{-0.6cm}
\end{table}

%
%-----------------------------------------------------------------------------------------------------------------------------
%
{}From the physical state viewpoint, solution 1 presented in Table IV corresponds to the excited states of $H_{x}$ combined with those of $H_{y}$. On the other hand, solutions 2 to 5, only valid for $p=0$, are respectively associated with the first excited state of $H_{x}$ and the ground state of $H_{y}$, the ground states of both $H_{x}$ and $H_{y}$, the ground state of $H_{x}$ and the first excited state of $H_{y}$, and the ground states of both $H_{x}$ and $H_{y}$ again. It is worth observing that the two solutions corresponding to the combination of the two ground states are actually characterized by the same structure functions when condition $p=0$ is taken into account.\par
%
%-------------------------------------------------------------------------------------------------------------------------------------
%
{}From Tables III and IV, we see that we do not obtain all the levels of the physical energy spectrum and that there also exist some additional degeneracies not described by the polynomial algebra. This phenonenon can be explained in terms of holes in the sequence of Hermite EOP.\par
%
%+++++++++++++++++++++++++++++++++++++++++++++++++++++++++++++++++++++++++++
%
\subsection{\boldmath Superintegrable systems from Laguerre EOP $L_{\alpha,m,n}^{\rm I}$, $L_{\alpha,m,n}^{\rm II}$, and $L_{\alpha,m,n}^{\rm III}$}

Let us now consider the case where $H_{x}$ corresponds to one of the systems associated with Laguerre EOP (Sec.~IIIB) and $H_{y}$ to a harmonic oscillator,
\begin{equation}
  H_{x}=-\frac{d^{2}}{dx^{2}} + \frac{1}{4}x^{2}+\frac{l(l+1)}{x^{2}}
  -2\left\{\frac{\dot{g}_m^{(\alpha)}}{g_{m}^{(\alpha)}}
  +2z \left[\frac{\ddot{g}_{m}^{(\alpha)}}{g_{m}^{(\alpha)}} 
  -\left(\frac{\dot{g}_{m}^{(\alpha)}}{g_{m}^{(\alpha)}}\right)^{2}\right]\right\}+C ,\label{sys3x}
\end{equation}
\begin{equation}
  H_{y}=-\frac{d^{2}}{dy^{2}} +  y^{2}, \label{sys3y}
\end{equation}
with integrals of motion given by Eq.~\eqref{KI}. The energy spectrum of the physical states for the three cases can be written as
\begin{equation}
  E_{\rm I}=2\nu_{x}+2\nu_{y}+l+\frac{3}{2},\quad l>0, \quad  \nu_{x},\nu_{y}=0,1,2,\ldots,\label{ener2dI}
\end{equation}
\begin{equation}
  E_{\rm II}=2\nu_{x}+2\nu_{y}+l+\frac{7}{2},\quad l>m-\frac{3}{2}, \quad  \nu_{x},\nu_{y}=0,1,2,\ldots,  
  \label{ener2dII}
\end{equation}
\begin{equation}
\begin{split}
  &E_{\rm III}=2\nu_{x}+2\nu_{y}+l+\frac{7}{2},\quad l>m-\frac{3}{2}, \quad m {\ \rm even}, \\ 
  &\nu_{x}=-m-1,0,1,\ldots, \quad \nu_{y}=0,1,2,\ldots.  
\end{split}  \label{ener2dIII}
\end{equation}
\par
%
%------------------------------------------------------------------------------------------------------------------------------------
%
{}From the ladder operators presented in Sec.~III, we obtain that the two PHA are given by Eqs.~\eqref{phax1}--\eqref{phay2} with $\lambda_{x}=\lambda_{y}=2$ and
\begin{equation}
\begin{split}
  Q(H_{x})&=\frac{1}{16}(2H_{x}-3-2l_{x}')(2H_{x}-1+2l_{x}')(H_{x}-2-E_{x})(H_{x}-E_{x}), \\
  S(H_{y})&=H_{y}-1,
  \end{split} \label{QS3}
\end{equation}
where $l'_x$ and $E_x$ assume different values according to whether the Laguerre EOP belong to Case I, II, or III (see Table II). The structure function, derived from~\eqref{functionF} and~\eqref{QS3}, takes the form
\begin{equation}
\begin{split}
  \Phi(E,u,x)&=\frac{1}{4} \left(\frac{E}{2}+2x+2u-\frac{3}{2}-l_{x}'\right) \left(\frac{E}{2}+2x+2u-
       \frac{1}{2}+l_{x}'\right) \\
  &\quad \times\left(\frac{E}{2}+2x+2u-2-E_{x}\right) \left(\frac{E}{2}+2x+2u-E_{x}\right) \\
  &\quad \times\left(\frac{E}{2}-2x-2u+1\right).  
\end{split}  \label{functionF3}
\end{equation}
\par
%
%------------------------------------------------------------------------------------------------------------------------------------
%
Equation~\eqref{constraints} leads to five solutions for the parameter $u$,
\begin{equation}
\begin{split}
  u_{1}&=-\frac{E}{4}+\frac{3}{4}+\frac{l_{x}'}{2},\quad u_{2}=-\frac{E}{4}+\frac{1}{4}-\frac{l_{x}'}{2}, \\
  u_{3}&=-\frac{E}{4}+1+\frac{E_{x}}{2},\quad u_{4}=-\frac{E}{4}+\frac{E_{x}}{2},\quad 
       u_{5}=\frac{E}{4}+\frac{1}{2}.
\end{split}  \label{paramu3}
\end{equation}
The finite-dimensional unitary representations are presented in Table V.

\begin{table}[h]
\begin{center}
\begin{tabular}{|l|l|l|l|l|l|l|}
\hline
  Case  & & $u$   &  $p$ &  Energy $E$  &  Structure function $\Phi$  &  Physical states   \\
\hline
  I & 1 & $u_{1}$  & $\mathbb{N}$  &  $2p+\frac{3}{2}+l$  & $x(p+1-x)(2x-1+2l)$ & $\nu_{x}=0,1,2,\ldots$,  \\
  &  &   &   &    & $\times (2x+2m+2l-1)(2x+2m+2l+1)$ & $\nu_{y}=0,1,2,\ldots$  \\ 
\hline
  II  & 1 & $u_{1}$   & $\mathbb{N}$  &  $2p+\frac{7}{2}+l$ & $x(p+1-x)(3+2l+2x)$ &   $\nu_{x}=0,1,2,\ldots$,   \\  
  &  &    &   &   &  $\times (1+2l-2m+2x)(3+2l-2m+2x)$   &  $\nu_{y}=0,1,2,\ldots$   \\  
\hline
  III  & 1 & $u_{1}$   & $\mathbb{N}$  &  $2p+\frac{7}{2}+l$ & $4x(p+1-x)(m+x)$ &   $\nu_{x}=0,1,2,\ldots$,   \\  
  &  &     &   &   & $\times (1+m+x)(3+2l+2x)$  & $\nu_{y}=0,1,2,...$  \\  
\hline
  III  & 2 & $u_{4}$   & 0  &  $2p+\frac{3}{2}+l-2m$ & $4x(p+1-x)(x-1)(x-1-m)$ &   $\nu_{x}=-m-1$,  \\ 
  & & & & & $\times (2x+1+2l-2m)$ & $\nu_{y}=0$ \\  
\hline
\end{tabular}
\caption{\label{tab:5/tc} Finite-dimensional unitary representations for the superintegrable system given by Eqs.~\eqref{sys3x} and~\eqref{sys3y} and related to Laguerre EOP.}
\end{center}
\vspace{-0.6cm}
\end{table}

%
%--------------------------------------------------------------------------------------------------------------------------------------
%
In Cases I and II, the solution is unique and provides all the energy levels with all their degeneracies. In Case III, we observe the same phenomenon as in Case 1 involving Hermite EOP. Solution 1 indeed corresponds to all excited states of $H_{x}$ combined with all states of $H_{y}$, while solution 2 is associated with the ground state of $H_{x}$ and that of $H_{y}$.\par
%
%++++++++++++++++++++++++++++++++++++++++++++++++++++++++++++++++++++++++++
%
\subsection{\boldmath Superintegrable system from Laguerre EOP $L_{\alpha,m_{1},m_{2},n}^{\rm I,II}$}

Let us finally consider a superintegrable system where $H_{x}$ is constructed from Laguerre EOP associated with second-order SUSYQM (Sec.~IIIC) and $H_{y}$ corresponds to a harmonic oscillator,
\begin{equation}
  H_{x}=-\frac{d^{2}}{dx^{2}} + \frac{1}{4}x^2 + \frac{l(l+1)}{x^2} 
  -2\left\{\frac{\dot{g}_{\mu}^{\alpha}}{g_{\mu}^{\alpha}}
  +2z\left[\frac{\ddot{g}_{\mu}^{\alpha}}{g_{\mu}^{\alpha}}
  -\left(\frac{\dot{g}_{\mu}^{\alpha}}{g_{\mu}^{\alpha}}\right)^{2}\right]\right\}, \label{sys4x}
\end{equation}
\begin{equation}
  H_{y}=-\frac{d^{2}}{dy^{2}} +  y^{2}. \label{sys4y}
\end{equation}
The integrals of motion are given by Eq.~\eqref{KI} again. The energy spectrum of the system can be written as
\begin{equation}
  E=E_{x}+E_{y}=2\nu_{x}+2\nu_{y}+l+\frac{5}{2}, \quad l > m_2-\frac{1}{2}, \quad \nu_{x},\nu_{y}=0,1,2,\ldots.  
  \label{ener4}
\end{equation}
\par
%
%------------------------------------------------------------------------------------------------------------------------------------
%
The PHA are given by Eqs.~\eqref{phax1}--\eqref{phay2} with $\lambda_{x}=\lambda_{y}=2$ and
\begin{equation}
\begin{split}
  Q(H_{x})&=\frac{1}{16}(2H_{x}-3-2l)(2H_{x}-1+2l)\left(H_{x}+l+2m_{1}-\frac{1}{2}\right) \\
  & \quad \times \left(H_{x}+l+2m_{1}+\frac{3}{2}\right)\left(H_{x}+l-2m_{2}-\frac{5}{2}\right)
      \left(H_{x}+l-2m_{2}-\frac{1}{2}\right), \\
  S(H_{y})&=H_{y}-1,
\end{split}  \label{QS4}
\end{equation}
leading to the structure function
\begin{equation}
\begin{split}
  \Phi(E,u,x)&=\frac{1}{4}\left(\frac{E}{2}+2x+2u-l-\frac{3}{2}\right)\left(\frac{E}{2}+2x+2u+l-\frac{1}{2}\right) \\
  & \quad \times \left(\frac{E}{2}+2x+2u+l+2m_{1}-\frac{1}{2}\right) \left(\frac{E}{2}+2x+2u+l+2m_{1}+\frac{3}{2}
       \right) \\
  & \quad \times \left(\frac{E}{2}+2x+2u+l-2m_{2}-\frac{5}{2}\right) \left(\frac{E}{2}+2x+2u+l-2m_{2}-\frac{1}{2}
       \right) \\
  & \quad \times \left(\frac{E}{2}-2x-2u+1\right).
\end{split}  \label{Function4}
\end{equation}
\par
%
%-----------------------------------------------------------------------------------------------------------------------------------
%
{}From~\eqref{constraints}, we obtain the following seven solutions for the parameter $u$,
\begin{equation}
\begin{split}
  u_{1}&=-\frac{E}{4}+\frac{l}{2}+\frac{3}{4},\quad u_{2}=-\frac{E}{4}-\frac{l}{2}+\frac{1}{4},\quad 
      u_{3}=-\frac{E}{4}-\frac{l}{2}-m_{1}+\frac{1}{4}, \\ 
  u_{4}&=-\frac{E}{4}-\frac{l}{2}-m_{1}-\frac{3}{4},\quad u_{5}=-\frac{E}{4}-\frac{l}{2}+m_{2}+\frac{5}{4},\quad 
      u_{6}=-\frac{E}{4}-\frac{l}{2}+m_{2}+\frac{1}{4}, \\
  u_{7}&=\frac{E}{4}+\frac{1}{2},
\end{split}  \label{paramu4}
\end{equation}
but only one finite-dimensional unitary representation presented in Table VI.

\begin{table}[h]
\begin{center}
\begin{tabular}{|l|l|l|l|l|l|}
\hline
  & $u$   &  $p$ &  Energy $E$ &  Structure function $\Phi$  &  Physical states   \\
\hline
  1 & $u_{1}$  & $\mathbb{N}$  &  $2p+l+\frac{5}{2}$  & $32x(p+1-x)(x+l+\frac{1}{2})(x+l+m_{1}+\frac{1}{2})
       $ & $\nu_{x}=0,1,2,\ldots$,   \\
   &   &   &    & $\times (x+l+m_{1}+\frac{3}{2})(x+l-m_{2}-\frac{1}{2})(x+l-m_{2}+\frac{1}{2})$ & $\nu_{y}
       =0,1,2,\ldots$ \\  
\hline
\end{tabular}
\caption{\label{tab:5/tc} Finite-dimensional unitary representation for the superintegrable system given by Eqs.~\eqref{sys4x} and~\eqref{sys4y} and related to Laguerre EOP.}
\end{center}
\vspace{-0.6cm}
\end{table}
\par
%
%-----------------------------------------------------------------------------------------------------------------------------------
%
In this case, this unique solution corresponds to all physical states.\par
%
%==========================================================================
%
\section{Conclusion}

In this paper, we introduced new superintegrable systems from SUSYQM and Hermite and Laguerre EOP. These systems possess higher-order integrals of motion that generate polynomial algebras. Furthermore, we obtained the finite-dimensional unitary representations of these polynomial algebras and presented them in Tables III--VI. Moreover, as two families introduced here include two systems previously obtained by Gravel \cite{Gra}, this paper also allows to understand the connections between them and EOP.\par
%
%---------------------------------------------------------------------------------------------------------------------------------
%
Many solutions presented in Tables III--V are only valid for $p=0$ and thus the energy spectrum is not recovered entirely from the finite-dimensional unitary representations. This phenomenon is associated with holes in the sequence of polynomials. As a consequence, $p+1$ is no longer the actual degeneracy of energy levels and thus the polynomial algebras do not describe all the degeneracies, as usually observed in the case of quadratically superintegrable systems.\par
%
%--------------------------------------------------------------------------------------------------------------------------------
%
The nature of these degeneracies seems to differ from that of the usual degeneracies obtained in the context of superintegrable systems (often called dynamical degeneracies) and explained by the polynomial algebra generated by the integrals of motion. This aspect needs further investigations and, in particular, the existence of a larger algebraic structure that would explain all the degeneracies is an interesting open question.\par
%
%-----------------------------------------------------------------------------------------------------------------------------
%
Quantum exactly solvable systems based on $k$-th order SUSYQM and connected with EOP can be introduced along the same lines as those presented in Sec.~III \cite{Que3}. They would allow the construction of superintegrable systems extending the results presented in this paper.\par
%
%=============================================================================
% 
\textbf{Acknowledgments} 

The research of I.M. was supported by the Australian Research Council through Discovery Project DP110101414. The article was written in part while he was visiting the Universit\'e Libre de Bruxelles. He thanks the National Fund for Scientific Research (FNRS), Belgium, for financial support during his stay.\par
%
%==========================================================================
%

\end{document}